\documentstyle[preprint,aps]{revtex}
\begin{document}
\draft
\title{\bf Scale Factor Duality and the Energy Condition Inequalities}
\author{Sayan Kar \thanks{Electronic Address :
sayan@iucaa.ernet.in}} 
\address{Inter University Centre for Astronomy and Astrophysics,\\
Post Bag 4, Ganeshkhind, Pune, 411 007, INDIA}
\maketitle
\parshape=1 0.75in 5.5in
\begin{abstract}
We demonstrate, by a simple analysis, that cosmological line elements
related by scale factor duality also exhibit 
a duality with respect to the conservation/violation of the
Weak Energy Condition (WEC) by the matter that acts as 
the source in the one-loop $\beta$ function equations for the
metric coupling written {\em explicitly} in
the form of the Einstein equations. Furthermore, a study of
specific pairs of line elements (obtained via O(d,d) transformations)
hints at a possible
generalisation of the above duality w.r.t. WEC for the 
case of $O(d,d)$ related spacetimes.
Consequences and extensions thereof are also pointed out. 
\end{abstract}
\vskip 0.125 in
\parshape=1 0.75in 5.5in

\newpage

Scale factor duality as a symmetry of the string equations of
motion in a curved background as well as low energy effective
string actions was first discovered by Veneziano {\cite{gaby:plb}}
in 1991. Shortly thereafter, it was extended to a more 
general symmetry
of the effective action known as $O(d,d)$ {\cite{mv:plb}}
where $d$ denotes the number of coordinates of which the 
metric and matter fields are independent.
Subsequent generalisations to background fields
depending on more than one coordinate have been
carried out in {\cite{as:plb}}. 
Such noncompact symmetries of string theory
{\cite{ms:npb}} have been exploited to a great extent
to construct inequivalent string vacua and has yielded
interesting background geometries representing black holes
{\cite{as:rev}} and black p--branes {\cite{as:plb}}, {\cite{gh:ictp}}
as well as cosmologies {\cite{ven:book}}, {\cite{gasp:book}}.

Stringy cosmologies (for reviews, see {\cite{ven:book}},
{\cite{gasp:book}} ) have certain characteristic features apart from
the symmetry of scale factor duality. Firstly, one does not need to
bring in an ad--hoc scalar field to get the necessary inflationary
phase. The dilaton field which arises as one of the massless
excitations of the string world sheet serves the purpose. Moreover,
there exists the notion of a phase termed as pre--big--bang
($t< 0$) during
which we have a rapidly inflating universe with a scale factor
generically obeying a pole law ( $a(t) = (-t)^{\beta}$ ; ($\beta < 0$
). This new phase is expected to end at $t=0$ where a FRW evolution
($a(t) = t^{\beta}$; $0 < \beta < 1$) takes over and we finally end
up with our universe today. Unfortunately, the transition from the
pre--big--bang epoch to the FRW phase is plagued by the presence
of a singularity, which, if absent would have solved the problem of
singularities in GR, at least, in a cosmological setting.
The generic presence of a singularity in the transition epoch which
spoils a smooth crossover has been termed as the graceful exit problem
in string cosmology {\cite{rv:plb}}. Recently, Rey{\cite{sjr:prl96}} has claimed
that by introducing quantum back reaction it is possible to
avoid the graceful exit problem at least within the limits of
a two dimensional model like that of CGHS {\cite{cghs:prd91}}.
Whether an extension of this to four dimensions is possible or
not is still an open question.
On the other hand, a quantum cosmology approach to graceful
exit has been advocated in {\cite{jmu:hepth}}.

The low energy effective theory that emerges out of string theory
is much like Einstein gravity. The equations are derivable from an
action which resembles that of a Brans--Dicke theory with the
$\omega$ parameter set to $-1$. Specifically, one has :

\begin{equation}
S_{eff} = \int d^{D}x \sqrt{-g} e^{-2\phi}\left [V -  R - 4(\nabla\phi)^{2}
+ \frac{1}{12} H_{\mu\nu\alpha}H^{\mu\nu\alpha} \right ]
\end{equation}

This is the bosonic sector of the genus--zero, low energy action for 
closed superstrings
in the limit when inverse string tension (or $\alpha^{\prime}$
) goes 
to zero. Here, $H_{\mu\nu\lambda}$ is the third--rank
antisymmetric tensor field, $\phi$ the
dilaton field and $V$ contains contributions from
the dilaton potential and the cosmological constant.
The $\beta$ function equations which are obtained by
imposing quantum conformal invariance in the worldsheet
sigma model can be derived from this
action by performing appropriate variations.

Obviously, there are two frames in which the metrics look very different--
the string (Brans--Dicke) and the Einstein frame. These frames are
related to each other by a conformal transformation. We shall
exclusively work in the string frame.

The Einstein--like equations can be written in the form $G_{\mu\nu}
= e^{2\phi}T_{\mu\nu}$ where $T_{\mu\nu} = T_{\mu\nu}^{\phi}
+ T_{\mu\nu}^{M}$ ($M$ denotes matter fields
other than the dilaton). These equations are exactly those for
Brans--Dicke theory with the parameter $\omega = -1$.
One might argue that it is not proper to rewrite the
$\beta$ function equations in an explicit Einstein form
because at the level of the worldsheet sigma model
$g_{\mu\nu}$, $\phi$ or $B_{\mu \nu}$ -- all have the
same status--they are the $background$ fields in which
strings propagate.
We give here two reasons justifying our stand regarding
rewriting the $\beta$ function equations as Einstein equations.

(i) If one has to compare and contrast low energy effective
string models with GR then it is essential to rewrite the
$\beta$ function equations in the form of a BD theory.
There is nothing erroneous in this -- no conservation laws
are violated and everything is consistent. The $\phi$
and $B_{\mu\nu}$ fields as well as others act as
sources for the metric. Moreover, it has been
claimed that the probable violation of the WEC essentially
invalidates the applicability of the singularity theorems
of GR to the case of string inspired gravity {\cite{gh:ictp}}. This is
ofcourse not entirely true as is shown for the
case of black holes in {\cite{sk:hepth}} and for cosmologies
in this paper.

(ii) At the level of pseudo -- Riemannian geometry
the Energy conditions can be thought of as conditions
on the Ricci tensor of the geometry evaluated along
a timeike/null geodesic, which ensures the occurence 
 of geodesic focusing.  Therefore, one can check the
 energy conditions (or, more precisely the null or
 timelike convergence condition) solely with the purpose of
 finding out whether geodesics can focus in a certain
 geometry. The presence of repulsive forces does actually
 emerge from this sort of an analysis.

We shall assume $T_{\mu\nu}^{M} = 0$
(although there do exist many solutions 
with axionic, moduli, other matter fields as well as
higher order terms in the Lagrangian {\cite{em:hepth}}).
Therefore we have only the dilaton field to worry about.
The stress-energy for the dilaton field is given as :

\begin{equation}
T_{\mu\nu}^{\phi} = e^{-2\phi} \left [ -2\nabla_{\mu}\nabla_{\nu} \phi
+ g_{\mu\nu} \nabla^{2} \phi \right ]
\end{equation}

One can therefore check the energy condition inequalities and 
conclude about the nature of the dilatonic matter that threads
a stringy solution. This exercise has been recently done for the
stringy black hole geometries {\cite{sk:hepth}}. We intend to
do the same for stringy cosmologies in this paper. As demonstrated
in a large part of the paper we find that scale--factor
duality is directly related to the violation/conservation of the
WEC. {\em If a scale factor $a(t)$ is generated out of matter
fields which satisfy the WEC then its dual ${\tilde a}(t) = \frac{1}{a(t)}$
can violate the WEC. } Thus a pre--big--bang phase can be marked
by a violation of the WEC whereas the FRW phase must necessarily satisfy
it. This feature is irrespective of whether graceful exit happens or not.
It should be mentioned that a duality in the values of the ADM masses
(in the sense of opposite signs) of pairs of T--dual solutions of 
low energy effective string
theory had been noted for special cases in 
{\cite{hw:prd}}. 
However, it was shown that there do exist counterexamples
to this and therefore  it is not a generic feature of T--dual 
spacetimes {\cite{hw:prd}}. Brief discussions on the violation/
conservation of the Energy
Conditions and their role in avoiding the singularity
theorems can be also found in {\cite{gmv:plb91}}, {\cite{art:npb94}}.   

To begin, let us write down the line element for a homogeneous,
isotropic cosmological model in $D$ spacetime dimensions.
This is given as :

\begin{equation}
ds^{2} = -dt^{2} +a^{2}(t)\left [ \frac{dr^{2}}{1-kr^{2}} +
r^{2}d\Omega_{D-2}^{2} \right ]
\end{equation}

where $a(t)$ is the scale factor and the term in square brackets
can be thought of as a metric on $S^{D-2}$, $R^{D-2}$ or $H^{D-2}$ (for $k=1, 
0, -1$ respectively). We shall work mostly with the $k=0$ case.
From the metric in (1) we can write down the Einstein tensor
$G_{\mu\nu}$ and hence equate it to the matter stress energy
$e^{2\phi}T_{\mu\nu}^{\phi}$.
Defining a diagonal $T_{\mu\nu}^{\phi}$ (which in actuality
is comprised out of the dilatonic matter) by the functions :

\begin{equation}
e^{2\phi}T_{00}^{\phi} = \rho (t) \quad ; \quad e^{2\phi}T_{ii}^{\phi} = p(t)
\end{equation}

we can straightaway write down the Einstein equations and the 
Weak Energy Condition (WEC) inequalities. Before we do that
let us recall the WEC .

{\em Weak Energy Condition :}

If $T_{\mu\nu} $ is the matter energy momentum tensor then 
the condition $T_{\mu\nu}\xi^{\mu}\xi^{\nu} \ge 0$ for all
nonspacelike $\xi^{\mu}$ is known as the WEC. For a diagonal
stress energy tensor ($T_{\mu\nu} \equiv diag (\rho, p_{i})$)
we have :

\begin{equation}
\rho \ge 0 \quad ; \quad \rho + p_{i} \ge 0 \quad (i=1,2,....D-2)
\end{equation}

Physically, the WEC implies the fact that matter energy density
has to be positive in all frames of reference. The $\rho + p_{i} \ge 0$
inequalities can be shown to be equivalent to the positivity of
energy density in a suitably chosen frame of reference ($\rho ^{\prime}
\ge 0$.
The subset of conditions obtained by assuming $\xi^{\mu}$ to be null
are known as the Null Energy Condition (NEC). This involves (for a
diagonal $T_{\mu\nu}$) only the $\rho + p_{i} \ge 0$ inequalities.
(In a geometric sense one can prefer calling these conditions on
matter as those on geometry by referring to them as the timelike
convergence and null convergence condition).

We therefore have :

{\em (I) Einstein equations }:

\begin{eqnarray}
\rho (t) =  \left [ \frac{(D-1)(D-2)}{2}\left 
\{ \left ( \frac{\dot a}{a} \right )^{2}
+ \frac{k}{a^{2}} \right \}\right ] \\ 
p(t) =  \left [\frac{(D-2)}{2} 
\left \{ -(D-3) \left (\frac{\dot a}{a} \right )^{2} 
- \frac{\ddot a}{a}- \frac{(D-3)k}{a^{2}}
\right \} \right ]
\end{eqnarray}

{\em (II) WEC Inequalities }:

\begin{eqnarray}
\rho (t) = \left [
\frac{(D-1)(D-2)}{2} \left \{ \left ( \frac{\dot a}{a} \right )^{2}
+ \frac{k}{a^{2}} \right \} \right ] \ge 0 \\
\rho (t) + p(t) = \left [(D-2) \left \{ 
\left ( \frac{\dot a}{a}\right )^{2}
- \frac{\ddot a}{a} + \frac{k}{a^{2}} \right \} \right ] \ge 0
\end{eqnarray}

The Einstein equations for $k=0$ are invariant under the following set of
transformations which comprise the notion of scale factor duality.

\begin{equation}
\bar a = \frac{1}{a} \quad ; \quad \bar \phi = \phi -Tr({\ln a})
\end{equation}

Note that by this $\rho$, $p$ go over to $\bar \rho$ and $\bar p$
respectively. The invariance of the Einstein equations  comes from 
including the transformation for the $\phi$ field in the actual
expressions for $\rho$ and $p$.

For $k = 0$ one can see that a condition for the conservation of the
WEC inequalities is given as :

\begin{equation}
F(t) = \left ( \frac{\dot a}{a} \right )^{2} - \frac{\ddot a}{a} \ge 0
\end{equation}

Now let us replace $a(t)$ by its dual scale factor $\bar a (t) = \frac{1}{a}$
Note now that the condition in terms of the quantity $\bar a$
turns out to be 

\begin{equation}
- \left ( \frac{\dot {\bar a}}{\bar a} \right )^{2} + \frac{\ddot {\bar a}}{a}
\ge 0
\end{equation}

which is {\em exactly opposite } to the condition for the original scale
factor $a(t)$.
Therefore, if $a(t)$ satisfies the WEC, then $\bar a(t)$ must necessarily
violate it. Therefore, in choosing specific transformations
to have invariance (of the Einstein equations) we have 
sacrificed the positivity of matter stress energy in any
frame. The stress--energy for the field $\phi$ does conserve the
WEC (for choices of $a$) but that for $\bar \phi$ does not.
The opposite of this statement is also true for different
choices of $a$.

One can also see this by looking at the explicit expressions for
the WEC in terms of the scalar field stress energy. The WEC
turns out to be given as :

\begin{equation}
\rho = - 2{\dot \phi}^{2} + 6\dot \phi \frac{\dot a}{a} \ge 0
\end{equation}

\begin{equation}
\rho + p = -2{\dot \phi}^{2} + 4\dot \phi \frac{\dot a}{a} \ge 0
\end{equation}

Note that the first of these equations remain invariant 
under the scale factor duality transformations for the 
scale factor and the scalar field. The second one picks up
an overall negative sign and is therefore consistent with
the expression obtained by considering the R.H.S. of the 
Einstein equation.

Let us demonstrate this fact with a pair of dual scale factors
that generically arise in string cosmology. We have 

\begin{equation}
a(t) = (t)^{\beta} \quad (0 <\beta < 1) \quad ; \quad \bar a(t)
=(-t)^{\beta} \quad (\beta <0)
\end{equation}

Note that the domain of the dual scale factor is $t\le 0$ while
for the other one it is $t\ge 0$.

Evaluating the expression for $F(t)$ one can very easily see that
the dual scale factor would violate the WEC while the other one would
necessarily conserve it.
Therefore, the inflationary (pre--big--bang ) epoch is born out
of dilatonic matter violating the WEC. Recall that in the usual inflationary
scenario the matter stress energy of the scalar field violates the
Strong energy condition ($T_{\mu\nu}\xi^{\mu}\xi^{\nu} - \frac{1}{2}
Tg_{\mu\nu}\xi^{\mu}\xi^{\nu} \ge 0)$.

We now focus our attention on a more general class of metrics
which are given as :

\begin{equation}
ds^{2} = -dt^{2} + a_{1}^{2}(t)dx_{1}^{2} + a_{2}^{2} (t)dx_{2}^{2}
+ a_{3}^{2} (t)dx_{3}^{2} 
\end{equation}

One can identify this class with the Kasner--type models
(with power law choices for the scale factors), except that we do not necessarily impose any extra restrictions on the
powers of $t$. 

Note that our analysis can be easily extended to higher  dimensions
with essentially no modifications except for the presence of certain
dimensionality factors.

The WEC inequalities for this case (from the L . H. S. of the Einstein
equations) turn out to be given as :

\begin{equation}
\rho = \frac{\dot a_{1}}{a_{1}} \frac{\dot a_{2}}{a_{2}} + \frac{\dot a_{2}}
{a_{2}} \frac{\dot a_{3}}{a_{3}} + \frac{\dot a_{3}}{a_{3}}\frac{\dot a_{1}}
{a_{1}} \ge 0
\end{equation}

\begin{equation}
\rho + p_{1} = -\frac{\ddot a_{2}}{a_{2}} - \frac{\ddot a_{3}}{a_{3}}
+ \frac{\dot a_{1}}{a_{1}} \frac{\dot a_{2}}{a_{2}} + \frac{\dot a_{1}}
{a_{1}}
\frac{
\dot a_{3}}{a_{3}} \ge 0
\end{equation}

\begin{equation}
\rho + p_{2} = -\frac{\ddot a_{1}}{a_{1}} - \frac{\ddot a_{3}}{a_{3}}
+ \frac{\dot a_{1}}{a_{1}} \frac{\dot a_{2}}{a_{2}} + \frac{\dot a_{2}}
{a_{2}} \frac{\dot a_{3}}{a_{3}} \ge 0
\end{equation}

\begin{equation}
\rho + p_{3} = -\frac{\ddot a_{1}}{a_{1}} - \frac{\ddot a_{2}}{a_{2}}
+ \frac{\dot a_{1}}{a_{1}}\frac{\dot a_{2}}{a_{2}} + \frac{\dot a_{2}}
{\a_{2}}\frac{\dot a_{3}}{a_{3}} \ge 0
\end{equation}

It is not entirely apparent from the above expressions whether
the dual scale factors violate/satisfy the WEC or not.
The fact is that we cannot make such a strong statement
about {\em all} scale factors as we did in the case
of the homogeneous isotropic model discussed previously.

However, the following restricted statement can be made.
Assume pairs of dual scale factors which obey conditions :
either $\dot a_{i} > 0$, $\ddot a_{i} > 0$ (inflationary
regime) or $\dot a_{i}>0$ , $\ddot a_{i} < 0$  (FRW regime).
Now it is clear from the inequalities that in the FRW regime
the WEC must necessarily be satisfied whereas in the 
inflationary epoch it {\em can} to violated.
More precisely, the second inequality, for example, implies that

\begin{equation}
\frac{\ddot a_{2}}{a_{2}} + \frac{\ddot a_{3}}{a_{3}}
\le \frac{\dot a_{1}}{a_{1}} \left ( \frac{\dot a_{2}}{a_{2}}
+ \frac{\dot a_{3}}{a_{3}}
\right )
\end{equation}

In the FRW--like epoch ,the L.H.S. of the above inequality is 
always negative while the
R.H.S. is necessarily positive -- hence there is no way in which the
inequality can be violated. On the other hand, in the inflationary
era, violations of the WEC are possible although
it is also true that it may be satisfied.

Let us now work out explicitly the inequalities for scale
factors which obey a pole law in the inflationary epoch and
a power (fractional or integral, but necessarily positive)law in the FRW
--like regime.

{\em Inflationary era :}

\begin{equation}
a_{1} (t) = (-t)^{-p} \quad ; \quad a_{2}(t) = (-t)^{-q} \quad ; \quad
a_{3} (t) = (-t)^{-r} \qquad t< 0\quad ; \quad 0< p,q,r < 1
\end{equation}

{\em FRW--like era} :

\begin{equation}
a_{1} (t) = t^{p} \quad ; \quad a_{2} (t) = t^{q} \quad ; \quad
a_{3} (t) = t^{r} \qquad t \ge 0 \quad ; \quad 0<p,q,r <1 
\end{equation}

The WEC inequalities in the two epochs translate into
the following :

{\em Inflationary Era}:

\begin{eqnarray}
pq + qr + rp \ge 0 \\
-q^{2} - r^{2} + (q+r) (p-1) \ge 0 \\
-p^{2} - r^{2} + (p+r) (q-1) \ge 0 \\
-p^{2} - q^{2} + (p+q) (r-1) \ge 0 
\end{eqnarray}

The first of these is trivially satisfied. The second, third
and fourth imply :

\begin{eqnarray}
p\ge \frac{q^{2} + r^{2}}{q+ r} + 1\\
q\ge \frac{p^{2} + r^{2}}{p+r} + 1\\
r\ge \frac{p^{2} + q^{2}}{p+q} + 1\\
\end{eqnarray}

Since $p,q,r$ are all less than 1 none of the above three can be
satisfied.

{\em FRW--like Era} :

In this regime we have the following inequalities :

\begin{eqnarray}
pq + qr + rp \ge 0 \\
-q^{2} - r^{2} + (q+r) (p+1) \ge 0\\
-p^{2} - r^{2} + (p+r) (q+1) \ge 0\\
-p^{2} - q^{2} + (p+q) (r+1) \ge 0\\
\end{eqnarray}

The second, third and fourth inequalities therefore imply :

\begin{eqnarray}
p\ge \frac{q^{2} + r^{2}}{q+r} - 1 \\
q\ge \frac{p^{2} + r^{2}}{p+r} - 1 \\
r\ge \frac{p^{2} + r^{2}}{p+q} - 1
\end{eqnarray}

These are always satisfied, now because $p,q,r > 0$
.Therefore, we have demonstrated the existence of dual
scale factors which necessarily satisfy/violate the WEC
while being solutions to the field equations of
the theory.

Futhermore, one can ask whether the results stated above remain
valid for general Brans--Dicke theories i.e. with the
parameter $\omega$ not restricted to $ \omega = -1$.
Cosmological solutions as well as a generalisation of the
scale factor duality symmetry has been found by Lidsey
{\cite{lid:hepth}}. We briefly discuss the analogs of the
above results now.

The SFD transformations now take the form :

\begin{equation}
\bar a = a^{\frac{2+3\omega}{4+3\omega}} exp ( - \frac{1+\omega}{
4+3\omega}\phi )
\end{equation}
\begin{equation}
\bar \phi = - \frac{3}{4+3\omega}\ln a - \frac{2+3\omega}{4+3\omega}
\phi
\end{equation}
with $\omega \neq - \frac{4}{3}$.

As before, we obtain two branches related by SFD. The explicit 
solutions are :
\begin{equation}
a^{\pm} = t^{p_{\pm}} \quad ; \quad e^{\phi} = t^{3p_{\pm} - 1}
\end{equation}
where
\begin{equation}
p_{\pm} = \frac{1}{4+3\omega} \left [ 1+ \omega \pm \left (
1 + \frac{2\omega}{3}\right )^{\frac{1}{2}} \right ]
\end{equation}

It is easy to check that the WEC inequalities will be
satisfied in both the $+$ and $-$ branches 
(this depends on the sign of $p_{\pm}$ and therefore on the
value of $\omega$, which we assume to be positive). Note
that for $\omega = -1$ the duality between violation and
conservation in the two branches appears. This is an
additional check on our previous calculations.

Finally, let us discuss more general metrics which have
off diagonal elements in their spatial part. We characterise
such metrics by a generic form given as :

\begin{equation}
ds^{2} = -dt^{2} + {\cal G}_{ik}{\cal G}_{kj} dx^{i}dx^{j}
\end{equation}

where ${\cal G}_{ij}$ is a $d\times d$ matrix.  
As before, one needs to write down the Energy Condition (WEC)
inequalities and explore the consequences regarding their status.
We shall confine ourselves to specific cases in $2+1$ 
dimensions. 

Let us first consider the solution given in {\cite{ak:mpl91}}
for the equations of motion emerging from the low energy action
with a term $V(\phi)$ containing the contribution of the 
dilaton potential as well as the cosmological constant.
We have the following form for the line element and
the dilaton field.

\begin{equation}
ds^{2} = -dt^{2} + g\tanh^{2}t dx_{1}^{2}
+ \left (\psi + \frac{1}{4}gb^{2}\tanh^{2}t\right )dx_{2}^{2}
+ bg\tanh^{2}t dx_{1}dx_{2}
\end{equation}

\begin{equation}
\phi = -\ln\cosh^{2}t + \phi_{0} 
\end{equation}

where $V=4$.
A little algebra will reveal that the Einstein tensor $G_{\mu\nu}$
 (in the one form basis $dt,{\sqrt \psi} dx_{2},
{\sqrt g}\tanh t (dx_{1} + \frac{1}{2}b dx_{2})$ ) for this metric would turn out to be :

\begin{equation}
G_{00} = G_{11} = G_{01}= G_{12}= 0  
\quad ; \quad G_{22} = 2sech^{2}t \quad; \quad
\end{equation}
Therefore, it is clear that the WEC, which involves
checking out the positivity of $G_{00}, G_{00} + G_{ii}\quad (i=1,2)$
,will be satisfied by this geometry.

Now, consider the dual geometry obtaining by acting on this metric
by an $O(2,2)$ transformation. This is done by defining a matrix
$M$ containing in general both the matrix ${\cal G}(t)$ (spatial part of the metric)
and ${\cal B}(t)$ (the antisymmetric tensor potential) in the following
combination.

\begin{equation}
M = \pmatrix{ {\cal G}^{-1} & -{\cal G}^{-1}{\cal B} \cr
-{\cal B}{\cal G}^{-1} & {\cal G} - {\cal B}{\cal G}^{-1}{\cal B}} \\
\end{equation}

The set of transformations which leave the action invariant are
given as :

\begin{equation}
M\rightarrow \Omega M\Omega^{T} \quad ; \quad  \Phi = \phi + \ln \det G 
\rightarrow \Phi
\end{equation}

where $\Omega$ is an element of the O(d,d) group ($\Omega \eta \Omega^{T}
= \eta$ , $\eta$ is the Minkowski metric).

The solution given above is for ${\cal B}=0$. We now choose
$\Omega$ to be :

\begin{equation}
\Omega = \pmatrix { \Pi & I - \Pi \cr
		   I -\Pi & \Pi \\}
\end{equation}
with $\Pi^{2} = \Pi$.
In $2+1$ dimensions, $\Pi$ can be chosen as :

\begin{equation}
\Pi = \pmatrix {0 & 0 \cr
		0 & 1 }
\end{equation}

The $O(2,2)$ transformed version of the metric now becomes :

\begin{equation}
ds^{2} = -dt^{2} + \frac{1}{g\tanh^{2}t}dx_{1}^{2} + \psi dx_{2}^{2} 
\end{equation}

The $G_{\mu\nu}$ for this metric is given as :
\begin{equation}
G_{00} = G_{11} = G_{12} = G_{01}= 0 
\quad ; \quad G_{22} = -\frac{2}{\sinh ^{2} t} 
\end{equation}

which clearly violates the WEC for all $t$ !

Alternatively , one can choose the matrix $\Pi$ to be :

\begin{equation}
\Pi = \pmatrix {1 & 0 \cr
                0 & 0 }
\end{equation}

The $O(2,2)$ transformed metric now takes the form :

\begin{equation}
ds^{2} = -dt^{2} + A^{2}(t)\left ( g\psi\tanh^{2} t dx_{1}^{2} + dx_{2}^{2}
\right )
\end{equation}

where $A^{2}(t) = \frac{1}{\psi +\frac{1}{4}gb_{1}^{2}\tanh^{2}t}$.
Note that the $O(2,2)$ transformation generates a nontrivial torsion
$\cal B$. We, however, do not need its explicit form in the
discusion below.
The WEC inequalities for this geometry are somewhat more
involved. We require the following to hold true :

\begin{equation}
A^{4} sech^{2}t \left [ \tanh^{2}{t}\left ( \frac{1}{8}g^{2}b_{1}^{4}
+ \frac{3}{4} gb_{1}^{2}\psi \right ) - \frac{1}{4}gb_{1}^{2} \psi
\right ] \le 0
\end{equation}

\begin{equation}
A^{4}sech^{2} t \left [ -\frac{1}{4}gb_{1}^{2}\left ( 2+ sech^{2}t 
\right )\psi - 2\psi^{2}\right ] \le 0
\end{equation}

The question of whether the above inequalities hold good can only
be answered by assuming the positivity/negativity of $\psi$
(other constants are taken as positive).

If $\psi$ is positive then the second inequality is trivially true
whereas the first one would require $\psi \leq -\frac{1}{4} g b_{1}^{2}
$-- which straightaway contradicts the  assumption $\psi > 0$.
On the other hand, if $\psi$ is negative one can  easily show
that the Lorentzian signature of the metric is not retained for all $t$. 

Therefore , the $O(2,2)$ transformation yields a metric which 
violates the WEC -- the choice of 
the value of $\psi$ does not play any role. Recall
that the original metric (on which $O(2,2)$ was applied 
) satisfied the WEC irrespective of the sign or values of
the quantity $\psi$.

An extension to metrics in $3+1$ or higher dimensions    
can be achieved by following the algorithm given in
{\cite{kkk:mpl91}}. The results concerning the
status of the WEC are analogous to the $
2+1$ dimensional case. We shall dwell upon this as
well as many other cases in a future article {\cite{sk:fut}}. 

To conclude, let us first summarize the results.

(i) The duality between conservation/violation of the WEC
by dilatonic or other matter fields for spacetime geometries
related by discrete SFD has been established. We have shown it
for diagonal metrics which are isotropic as well as cases
where anisotropy is present (Kasner like models). A brief
statement about the continuation of these results to BD theories 
with $\omega \neq -1$ has also been made.

(ii) For the case of $O(d,d)$ related metrics we have outlined
special cases in $2+1$ dimensions. A similar duality (though
only through special cases)
with regard to the violation/ conservation of the WEC is 
obtained for pairs of $O(d,d)$ related geometries.

It would be worth obtaining a general statement regarding this
duality by studying the WEC inequalities for dual metrics
related by $O(d,d)$. 
This would basically imply a classification of solutions 
in terms of their conservation/violation of the WEC or
equivalently their features w.r.t. geodesic
focusing. 
Such a phenomenon seems to be a feature 
of the low energy string effective actions in the string frame.
Symmetries of the action and the equations of motion 
are (and should be) reflected on the behaviour and interplay of
matter and geometry. 

In the context of string cosmology, it is important to note
that a pre--big--bang phase must necessarily be born out
of matter violating the WEC. lt would be instructive to
check out the averaged energy conditions (AWEC, ANEC)
{\cite{ft:avg}} for this phase. Matter fields which generate
a pre--big--bang phase could therefore be quantum in nature,
considering the fact that quantum stress tensors can in
principle violate the local energy conditions {\cite{egy:il}} but may
satisfy their global (averaged) versions. 

The author thanks M. Gasperini, A. Kumar, J. Maharana and
T. Padmanabhan
for useful discussions. Financial support from the
Inter University Centre for Astronomy and
Astrophysics, Pune, India is also gratefully
acknowledged.

\end{document}